\documentclass[a4paper,twoside]{article}
\usepackage{amsmath,amssymb,amsfonts}
\pdfoutput=1
\usepackage{hyperref}
\usepackage{graphics}

\begin{document}
\title{Information geometry and entropy in a stochastic epidemic rate process}
\author{ C.T.J. Dodson \\{\small\it School of Mathematics, University of Manchester,
  Manchester M13 9PL, UK}\\
  {\small\it ctdodson@manchester.ac.uk}}
\pagestyle{myheadings}
\markboth{Information geometry and entropy in a stochastic epidemic rate process}{C.T.J. Dodson}
\date{}
\maketitle

\begin{abstract}
Epidemic models with inhomogeneous populations have been used to study major
outbreaks and recently Britton and Lindenstrand~\cite{BL} described the case
when latency and infectivity have independent gamma distributions. They found that
variability in these random variables had opposite effects on the epidemic
growth rate. That rate increased with greater variability in latency but decreased
with greater variability in infectivity. Here we extend their result by using the McKay bivariate
gamma distribution for the joint distribution of latency and infectivity, recovering
the above effects of variability but allowing possible correlation. We use methods
of stochastic rate processes to obtain explicit solutions for the growth of
the epidemic and the evolution of the inhomogeneity and information entropy.
We  obtain a closed
analytic solution to the evolution of the distribution of the number of uninfected
individuals as the epidemic proceeds, and a concomitant
expression for the decay of entropy.
The family of McKay bivariate gamma distributions has a tractable information
geometry which provides a framework
in which the evolution of distributions can be studied as the outbreak grows,
with a natural distance structure for quantitative tracking of progress.\\
{\bf Keywords:} Epidemic model, stochastic rate process, inhomogeneous, bivariate gamma,
information geometry, entropy, distribution evolution.
\end{abstract}
\section{Introduction}
Epidemiology is a big subject with a long history and a large literature;
standard texts on modeling include~\cite{bailey},~\cite{AM},~\cite{DH}, note also
the new volume on numerical methods~\cite{AB}.

The spreading of an infectious disease involves a (large) population in which an initially
small number of individuals are infected. Each infected individual is for a period of latency $L$
not yet infectious but at the end of the latent period the individual becomes infectious for
a period $I;$ models provide distributions for the random variables $L$ and $I.$
The rest of the population is susceptible to infection from infectious individuals;
this susceptibility is often taken to be a constant but later we shall consider a
population with an evolving inhomogeneous distribution of susceptibilities to infection.
An infectious individual has random infectious contacts at a rate $\lambda$ and
contact with a susceptible individual results in infection and then the
latent period of that individual commences. The so-called basic reproduction number is
$R_0=\lambda\mu_I,$ the product of the rate of infectious contacts and the
mean period of infectiousness $\mu_I.$

\cite{lloyd} discussed the sensitivity of dynamical properties
of an epidemic model to the choices of formulation and made use of a gamma distribution
for the period of infectiousness and allowed for optional seasonality.
\cite{CF} introduced a new approach to the analysis of epidemic
time series data to take account of partial observation of latency and the temporal
aggregation of observed data. They showed that homogeneous standard models can miss
key features of epidemics in large populations. Also, \cite{NCHW} devised
an estimate of reproduction number in terms of coarsely reported epidemic data,
showing that an ideal reporting interval is the mean generation time rather than
a fixed chronological interval. See also recent work by \cite{HIK} for related results
on partially observed data and by \cite{miller09} on general distributions of generating
 intervals.

\cite{Chowell} have edited a new collection of articles on
mathematical and statistical approaches
to epidemic modelling and Chapter 2 there, by G. Chowell and F. Bauer, gives a detailed study
of the basic reproduction rate in a variety of epidemic models. \cite{WL}
addressed the sensitivity of the reproduction number to the shape
of the distribution of generation intervals and obtained upper bounds even in the situation
of no information on shape.

Recently~\cite{BL} described a model
where the period of latency $L$ and the period of infectiousness $I$ have
independent gamma distributions. They found
that variability in these random variables had opposite effects on the epidemic
growth rate. That rate increased with greater variability in $L$ but decreased
with greater variability in $I.$ Here we extend their result by using the
McKay bivariate gamma distribution for the joint distribution of $L$ and $I,$
recovering the above effects of variability but allowing in case it may
be of relevance the possibility of correlation. One might imagine that in the case
of a disease in which the physical changes during latency lead to longer
future infectiousness if the period of their development is longer, then the random
variables $L$ and $I$ may have a positive correlation.  We
use methods of stochastic rate processes to obtain explicit solutions for the
growth of the epidemic and the evolution of the inhomogeneity and information
entropy. This admits a closed analytic solution to the evolution of the distribution
of the number of uninfected individuals as the epidemic proceeds, and a concomitant
expression for the decay of entropy. The family of McKay bivariate gamma
distributions has a tractable information geometry which provides a framework
in which the evolution of distributions can be studied as the outbreak grows,
with a natural distance structure for quantitative tracking of progress.

\section{Inhomogeneous Malthusian epidemic models}
In their discussion of epidemic modelling, \cite{BL} highlighted aspects
when stochastic features are more important than deterministic ones. In particular, they
described the importance of admitting random variables to represent the period of latency $L$ and the
period of infectiousness $I,$
Their standard susceptible-exposed-infectious-removed (SEIR) epidemic
model was elaborated using independent gamma distributions for  $L$ and
 $I$ with means $\mu_L,\mu_I$ and standard deviations $\sigma_L,\sigma_I.$
 The basic (mean) reproduction number is given by
\begin{equation}\label{R0}
    R_0= \lambda \mu_I
\end{equation}
where $\lambda$ is the rate of infectious contacts and $\mu_I$ is the mean length
of infectious period. An epidemic becomes a major outbreak if $R_0>1$
and then the number infected increases exponentially,
\begin{equation}\label{expinc}
n_I(t)\sim e^{r t}
\end{equation}
where the Malthusian parameter $r$ satisfies the equation
\begin{equation}\label{malth}
    E\left(e^{- r t} \lambda \ Prob\{L<t<L+I\}\right)= 1.
\end{equation}
Their independent bivariate model
expresses $\frac{r}{R_0}$ in terms of the parameters of the two gamma
distributions. They used means, $\mu_L,\mu_I$ and coefficients of variation
$\tau_L=\frac{\sigma_L}{\mu_L},\tau_I=\frac{\sigma_I}{\mu_I}$ to deduce
\begin{equation}\label{rate}
r=\frac{R_0}{\mu_I} \left(1+r \tau_L^2\mu_L \right)^{-1/\tau_L^2}
           \left(1- \left(1+r \tau_I^2\mu_I\right)^{-1/\tau_I^2} \right).
\end{equation}
Then \cite{BL} found from numerical analysis of (\ref{rate})
that, at fixed $R_0,$ the growth rate $r$ is monotonically
decreasing with $\mu_L,\mu_I$ and $\tau_I,$ but it is increasing with $\tau_L.$ So
increased variability in latency period increases the epidemic growth rate whereas
increased variability in infectious period decreases the epidemic growth rate.

\section{Bivariate gamma distribution of periods of latency and infectiousness}
The model described here adds to the work of \cite{BL}
in which they used {\em independent} univariate gamma distributions for the periods
of latency and infectiousness in an epidemic model that they illustrated
with data from the SARS outbreak~\cite{who}. They used numerical methods to obtain
approximate solutions. Our contribution is to
use a bivariate gamma distribution which allows positive correlation
between the random variables representing the periods of
latency and infectiousness. That could represent a situation where
physical changes during the latency period lead to longer
future infectiousness if the period of their development is longer.
We obtain a closed
analytic solution and show that the same qualitative
features persist in the presence of such correlation.
This makes available the analytic information geometry of the space of probability
densities, allowing comparison of possible trajectories for the epidemic
against, for example, exponential distributions for periods of
infectiousness or of latency, corresponding to underlying Poisson processes.

\begin{figure}
\begin{center}
\begin{picture}(300,150)(0,0)
\put(-50,-10){\resizebox{20cm}{!}{\includegraphics{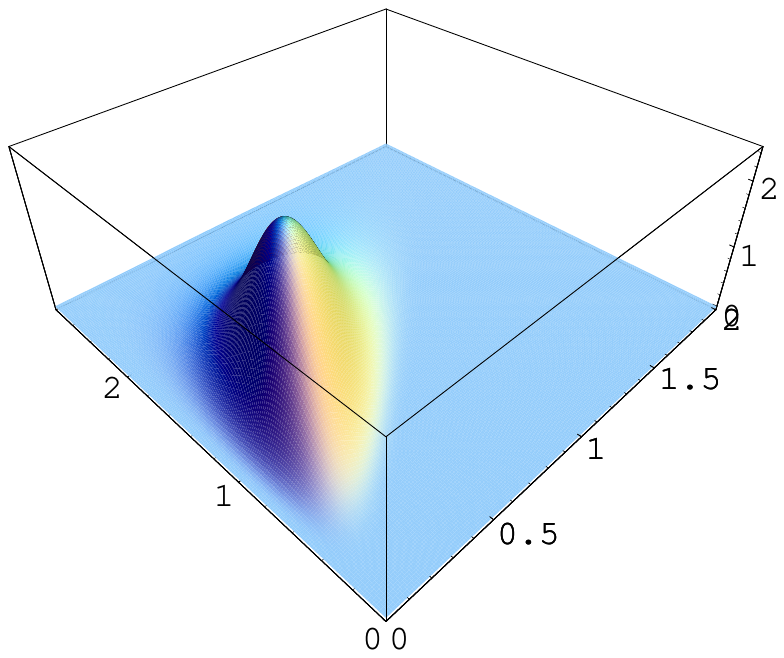}}}
\end{picture}
\end{center} \caption{{\em Part of the family of McKay bivariate
gamma probability density functions $f(x,y)$ with correlation
coefficient $\rho=0.6$ and $\alpha_1=5.$}}
\label{McKaypdf}
\end{figure}

Somewhat surprisingly, it is rather difficult to devise bivariate versions of Poisson, exponential
distributions or more generally gamma distributions that have reasonably simple form,
 and indeed only Freund bivariate exponential and
McKay bivariate gamma distributions seem to have tractable information geometry~\cite{InfoGeom}.
The family of McKay bivariate gamma density functions~
is defined on $ 0<x<y<\infty $ with parameters $
\alpha_{1},\sigma_{12},\alpha_{2}>0$ and probability density functions, Figure~\ref{McKaypdf},
\begin{eqnarray}
f(x,y;\alpha_{1},\sigma_{12},\alpha_{2}) =
\frac{(\frac{\alpha_{1}}{\sigma_{12}})^{\frac{(\alpha_{1}+\alpha_{2})}{2}}x^
{\alpha_{1}-1}(y-x)^{\alpha_{2}-1}
e^{-\sqrt{\frac{\alpha_{1}}{\sigma_{12}}}y}}{\Gamma(\alpha_{1})\Gamma(\alpha_
{2})} \ . \label{mckaydistribution4}
\end{eqnarray}
Here $\sigma_{12},$ which must be positive, is the covariance of $x$ and
$y$ and  $f(x,y)$ is the probability density for the two
 random variables $x$ and $y=x+z$ where $x$ and $z$ both have gamma density functions.

\begin{figure}
\begin{center}
\begin{picture}(300,170)(0,0)
\put(0,0){\resizebox{8 cm}{!}{\includegraphics{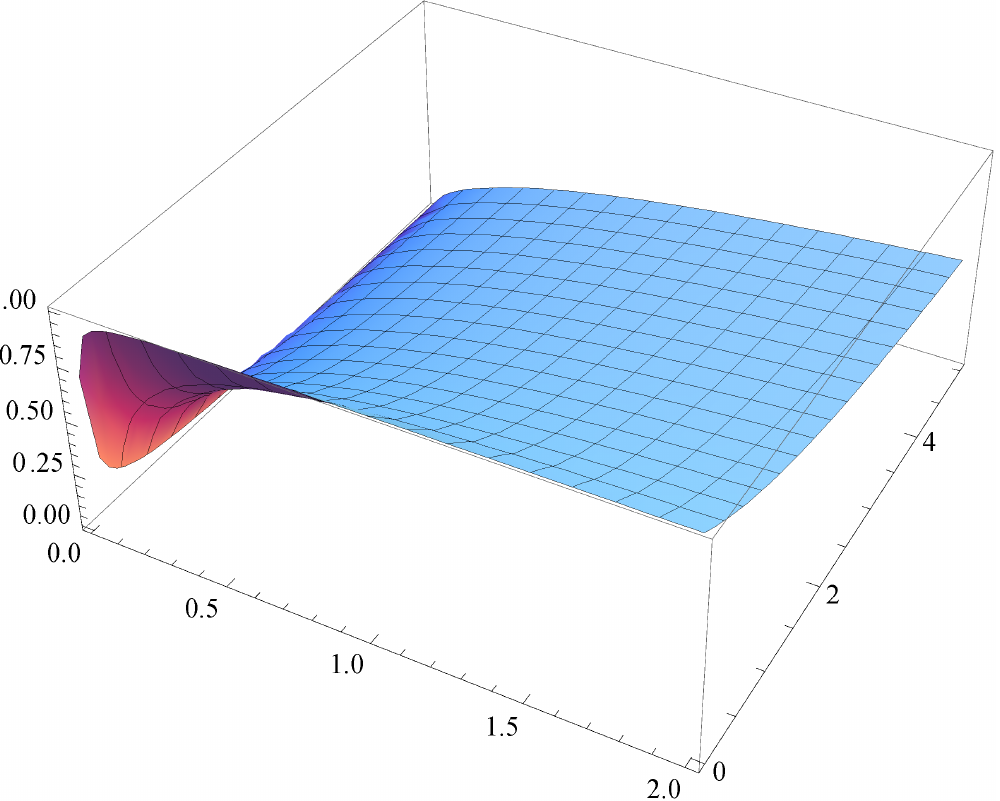}}}
\end{picture}
\end{center} \caption{{\em Correlation coefficient $\rho$
from equation (\ref{rhoa1a2}) for McKay probablity density
functions, in terms of $\alpha_1,\alpha_2.$ This represents
correlation between lengths of periods of latency and infectiousness.  }}
 \label{McKayrhoa1a2}
\end{figure}
 We obtain the means, standard deviations and coefficients of
 variation by direct integration:
 \begin{eqnarray}
   {\rm Means:}&& \ \mu_x = \sqrt{\alpha_1\sigma_{12}}, \ \
   \mu_y=\frac{(\alpha_1+\alpha_2)\sqrt{\sigma_{12}}}{\sqrt{\alpha_1}}, \  \
   \mu_z= \frac{\alpha_2\sqrt{\sigma_{12}}}{\sqrt{\alpha_1}}\\
   {\rm SDs:}&& \ \sigma_x = \sqrt{\sigma_{12}}, \ \
    \sigma_y = \sqrt{\frac{\sigma_{12}(\alpha_1+\alpha_2)}{\alpha_1 }}, \ \
    \sigma_z = \sqrt{\frac{\alpha_2\sigma_{12}}{\alpha_1}}\\
    {\rm CVs:}&& \ \tau_x  = \frac{1}{\sqrt{\alpha_1}}, \ \
    \tau_y= \frac{1}{\sqrt{\alpha_1+\alpha_2}}, \ \
    \tau_z=\frac{1}{\sqrt{\alpha_2}}
 \end{eqnarray}

The correlation coefficient,
and marginal probability density functions of $x$ and $y$ are
given by
\begin{eqnarray}
  \rho&=& \sqrt{\frac{\alpha_{1}}{\alpha_{1}+\alpha_{2}}} > 0 \label{rhoa1a2}\\
 f_{1}(x) &= & \frac{(\frac{\alpha_{1}}{\sigma_{12}})^{\frac{\alpha_{1}}{2}} x^
{\alpha_{1}-1}e^{-\sqrt{\frac{\alpha_{1}}{\sigma_{12}}}
x}}{\Gamma(\alpha_{1})},\quad x
>0 \\
 f_{2}(y)&=&\frac{(\frac{\alpha_{1}}{\sigma_{12}})^{\frac{(\alpha_{1}+\alpha_{2})}{2}}
y^{(\alpha_{1}+\alpha_{2})-1}e^
{-\sqrt{\frac{\alpha_{1}}{\sigma_{12}}}
y}}{\Gamma(\alpha_{1}+\alpha_{2})}, \quad y>0
\end{eqnarray}
Figure~\ref{McKayrhoa1a2} shows a plot of the correlation
coefficient from equation (\ref{rhoa1a2}). The marginal probability density
functions of latency period $x$ and infectiousness period $y$ are gamma with shape parameters
$\alpha_{1}$ and $\alpha_{1}+\alpha_{2}$, respectively.
It is not possible to choose parameters such that both marginal
 functions are exponential, so the two random variables cannot
 both arise from Poisson processes in this model.

\section{Stochastic rate processes}\label{SRP}
For a detailed monograph on stochastic epidemic models see \cite{AB}.
We consider here a class of simple stochastic rate processes
where a population $N,$ of uninfected individuals,
is classified by a smooth family of time-dependent probability density functions
$\{P_t, t\geq 0\}$ with random variable $a> 0,$  having at time $t$ mean $E_t(a)$ and variance
$\sigma^2(t).$
This situation was formulated by~\cite{Karev03},~\cite{Karev10} in the following way.
Let $l_t(a)$ represent the frequency at the $a$-cohort, then we have
\begin{eqnarray}
   N(t)&=& \int_0^\infty l_t(a) \, da  \ \ \ {\rm and} \ \ P_t(a) =\frac{l_t(a)}{N(t)}\label{rateproc} \\
                         \frac{dl_t(a)}{dt}&=&  -a l_t(a)  \ \ \ {\rm so} \ \ l_t(a) = l_0(a) e^{-at}
                         \end{eqnarray}
General solutions for these equations were given in \cite{Karev03}, from which we obtain
\begin{eqnarray}
                         N(t) &=& N(0)L_0(t) \ \ {\rm where} \
                         L_0(t)= \int_0^\infty P_0(a) e^{-at} \, da \label{L0}\\
                         \frac{dN}{dt}&=& -E_t(a) \, N \ \ {\rm where} \
                         E_t(a)= \int_0^\infty a \, P_t(a) \, da  = -\frac{d\log L_0}{dt} \label{Eta}\\
                          \frac{dE_t(a)}{dt}&=& -\sigma_t^2(a) = (E_t(a))^2 - E_t(a^2) \label{Vta}\\
      P_t(a) &=& e^{-at}\frac{P_0(a)}{L_0(t)} \ \ {\rm and} \ l_t(a) = e^{-at}L_0(t) \label{Pt}\\
                        \frac{dP_t(a)}{dt}&=&  P_t(a) (E_t(a)-a). \label{Prate}
                       \end{eqnarray}
Here $L_0(t)$ is the Laplace transform of
the initial probability density function $P_0(a)$
and so conversely $P_0(a)$ is the
inverse Laplace transform of the population (monotonic) decay solution $\frac{N(t)}{N(0)}.$ See
 \cite{Feller} for more discussion of the existence and uniqueness properties of the
correspondence between probability densities and their Laplace transforms.
In this section we shall use $N(t)$ to represent the decreasing population of
{\em uninfected} individuals as an epidemic grows.
 In our context of an epidemic model we might view the random
variable $a$ as a  feature representing susceptibility to infection in the population;
in general this distribution will evolve during the epidemic.
The model can be reformulated for a vector $(N^i(t))$
representing a composite population with a vector of distributions $(P_t^i)$ and a matrix of
variables $[a_{ij}]$.

\begin{figure}
\begin{center}
\begin{picture}(300,160)(0,0)
\put(-20,-20){\resizebox{8cm}{!}{\includegraphics{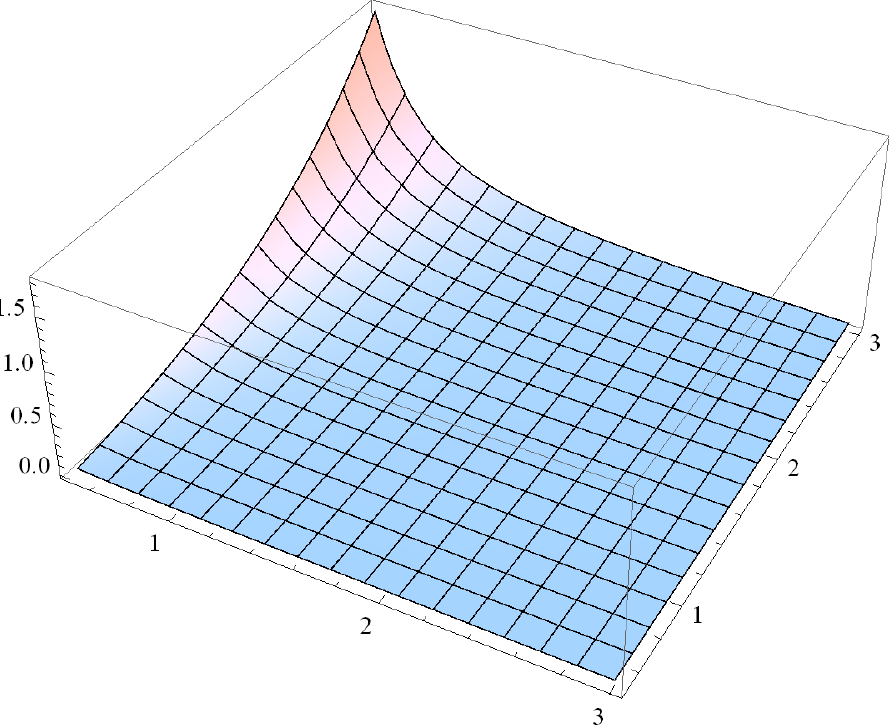}}}
\end{picture}
\end{center} \caption{{\em Plot of the Malthusian parameter $r$
against the coefficients of variation of periods of latency $\tau_x$ and infectiousness $\tau_y$
with means $\mu_x=3, \, \mu_y=2$ and $R_0=2.2$ from \cite{BL} for the SARS data~\cite{who}. So the
exponential growth rate of infection decreases with variability in latency period ($\tau_x$) but
increases with variability in infectiousness period ($\tau_y$). }}
\label{r123}
\end{figure}
\begin{figure}
\begin{center}
\begin{picture}(300,180)(0,0)
\put(0,-20){\resizebox{10cm}{!}{\includegraphics{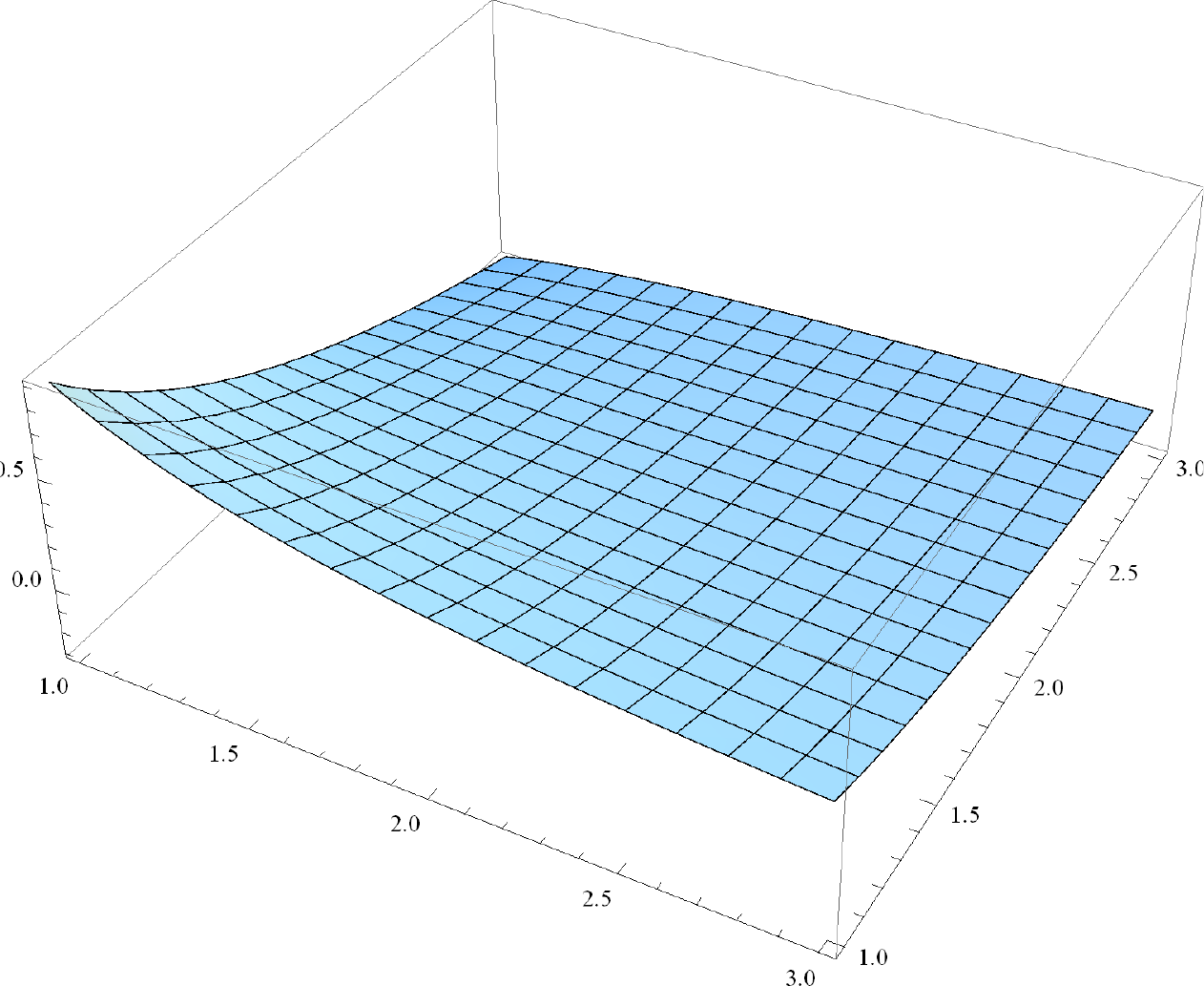}}}
\end{picture}
\end{center} \caption{{\em Plot of the Malthusian parameter $r$
against the means of periods of latency $\mu_x$ and infectiousness $\mu_y$
with coefficients of variation $\tau_x=\tau_y=4/7$ and $R_0=2.2$ from \cite{BL}
for the SARS data~\cite{who}.
 So the exponential growth rate of infection decreases with mean latency period ($\mu_x$) and
 with mean infectiousness period ($\mu_y$).}}
\label{r3}
\end{figure}
It is easy to deduce the rate process for entropy from Karev's model.
The Shannon entropy at time $t$ is
\begin{equation}\label{Sdef}
    S_t =  - E_t\left(\log P_t(a)\right) =
    - E_t\left(\log \frac{P_0(a) e^{-at}}{L_0(t)} \right)
\end{equation}
which reduces to
\begin{equation}\label{S}
    S_t = S_0 + \log L_0(t) + E_t(a) \, t .
\end{equation}
By using $\dot{E_t}(a) = -\sigma^2(t),$ the decay rate is then
\begin{equation}\label{Sdecay}
    \frac{dS_t}{dt} = - t \ \sigma^2(t) .
\end{equation}
This result shows how the variance controls the entropy change during
quite general inhomogeneous population processes.
In fact equation~(\ref{Sdecay}) and further related results were given also in
subsequent papers~\cite{Karev09},~\cite{Karev10}.
We note that the reverse process of population growth may have applications in constrained
disordering type situations~\cite{entflow}.

\begin{figure}
\begin{center}
\begin{picture}(300,150)(0,0)
\put(60,-10){\resizebox{7cm}{!}{\includegraphics{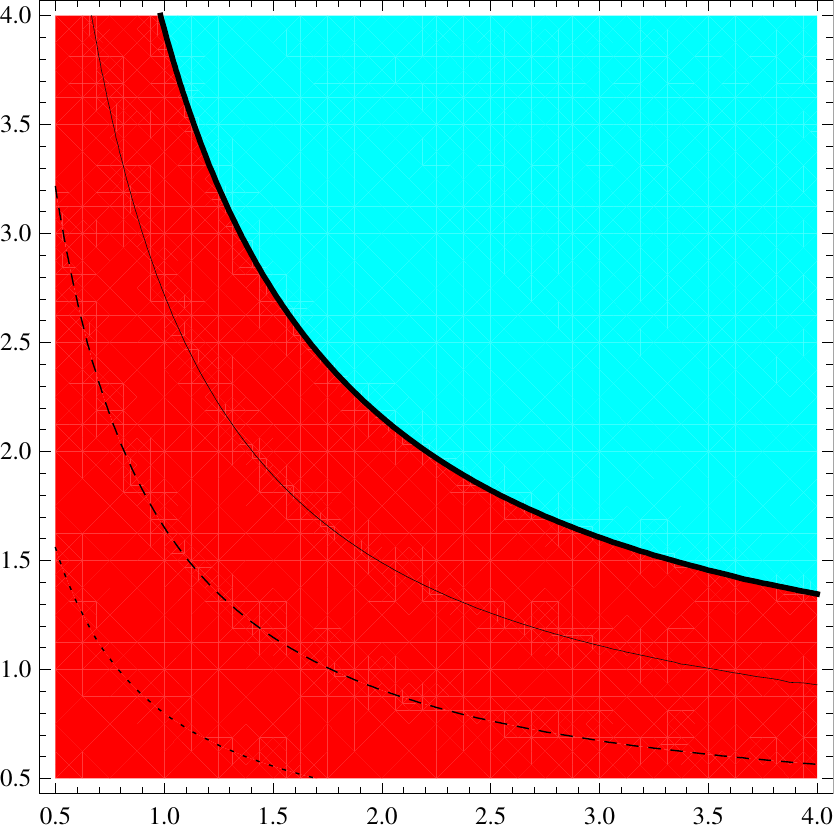}}}
\end{picture}
\end{center} \caption{{\em Contour plot of the contact rate $\lambda$
against the coefficients of variation of latency
period $\tau_x$ and of infectiousness period $\tau_y$
with mean latency period $\mu_x=3$ and $r=0.053$ from \cite{BL} for the SARS data~\cite{who}.
The levels are $\lambda=1.1$ dotted, $\lambda=1.5$ dashed, $\lambda=3$ thin
and $\lambda=10$ thick.}}
\label{l123}
\end{figure}

\begin{figure}
\begin{center}
\begin{picture}(300,190)(0,0)
\put(60,-10){\resizebox{7cm}{!}{\includegraphics{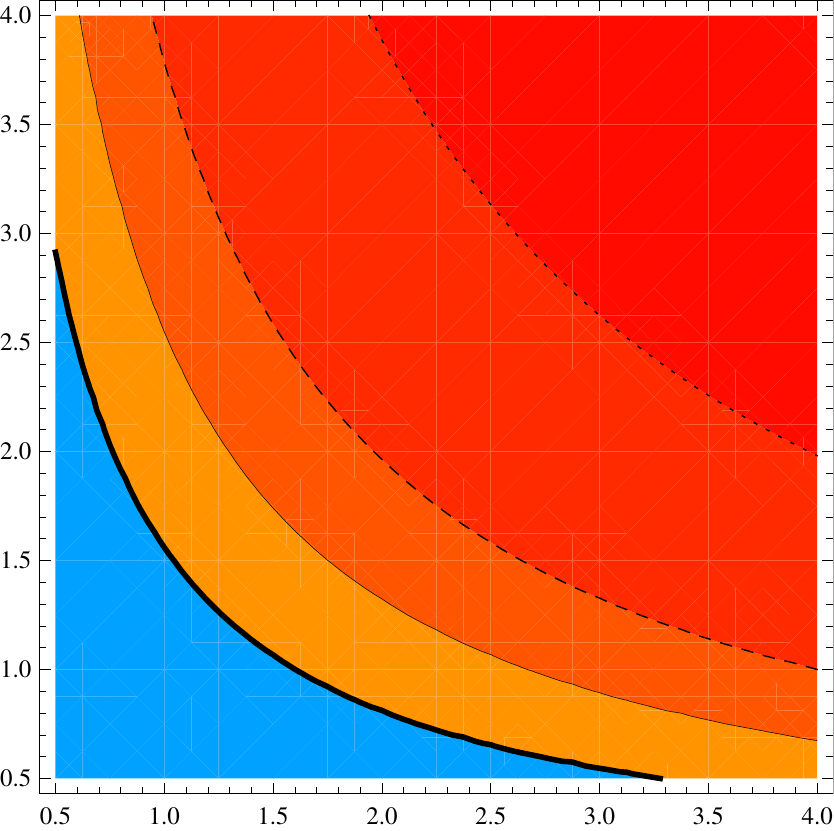}}}
\end{picture}
\end{center} \caption{{\em Contour plot of the contact rate $\lambda$
against the McKay parameters $\alpha_1$ and  $\alpha_2$
with mean latency period $\mu_x=3$ and $r=0.053$ from \cite{BL} for the SARS data~\cite{who}.
The levels are $\lambda=1.02$ dotted, $\lambda=1.04$ dashed, $\lambda=1.06$ thin
and $\lambda=1.1$ thick.}}
\label{l3}
\end{figure}

\subsection{Initial growth rate}
We follow the method of \cite{BL} in their \S3.2, to compute the
initial exponential growth rate of the epidemic from equation (\ref{malth}) which we write
for bivariate $x,y$ in the form
\begin{equation}\label{M}
    \int_0^\infty \int_y^x e^{-r(y-x)} \, \lambda \, f(x,y)\ dy \, dx = 1
\end{equation}
Here, from~\cite{BL},  $R_0=\lambda \mu_x$ for the average number of infections
per infective, so $\lambda$ is the contact rate; this gives the Malthusian parameter
analytically in explicit form as
\begin{equation}\label{r}
    r=\frac{1}{\mu_x \tau_x^2} \left( \left( \frac{R_0}{\mu_y}\right)^{\tau_y^2} -1\right).
\end{equation}
Thus,  $r$ is monotonically decreasing with $\mu_x, \, \mu_y$ and $\tau_x$ but increasing
with $\tau_y;$ Figure~\ref{r123} and Figure~\ref{r3} plot
typical values from the SARS epidemic~\cite{who} as used by  \cite{BL}.
Figure~\ref{l123} and Figure~\ref{l3}
show corresponding contour plots of the infectivity rate $\lambda.$
 So the bivariate gamma model reveals that the result of ~\cite{BL} for the dependence
of growth rate on variability in the periods of latency and infectiousness in the independent case
persists also in the presence of correlation between these two random variables.
Such a correlation may be relevant in particular applications, when physical changes evolve
during the latent period and influence the length of the subsequent infectiousness period.

We can estimate also the evolution of an inhomogeneous distribution
of susceptibility $a,$
 as the population $N(t)$ of uninfected individuals declines with time $t.$
For example, the case when the initial distribution $P_0(a)$ is a
gamma distribution with parameters $s,k$ was solved in \cite{Karev03} giving the result
\begin{equation}\label{gamevol}
    P_t(a) = \frac{P_0(a)}{L_0(t)} e^{-at} = \frac{(s+t)^k \, a^{k-1}}{\Gamma(k)}\, e^{-a(s+t)} ,
   \ \ \ \rm{for \ time} \ t\geq 0.
\end{equation}
Then the time dependences of mean, standard deviation and coefficient of variation are given by
\begin{eqnarray}
  \mu_a(t) &=& \frac{k}{s+t} \\
  \sigma_a(t) &=& \frac{\sqrt{k}}{s+t} \label{sz}\\
  \tau_a(t) &=& \frac{1}{\sqrt{k}}.
\end{eqnarray}
From (\ref{Sdecay}), we can see that the {\em rate} of entropy decrease
is greater for more variability in susceptibility.

\section{Information geometry of the space of McKay bivariate gamma distributions}
Information geometry of the smooth family $M$ of McKay bivariate
gamma probability density functions, which is of exponential type,
has been studied in detail in ~\cite{InfoGeom} Chapter 4. This provides a Riemannian
metric on $M,$  yielding a curved 3-manifold so the affine immersion
is a 3-dimensional object in $\mathbb{R}^4,$ which we can only represent in $\mathbb{R}^3$
through its 2-dimensional submanifolds. Here we illustrate how the geometry may
nevertheless be used to provide a natural distance structure on the
space of the McKay distributions used in our epidemic model.

First we measure distances from distributions with exponential marginal
distributions---those for which  $\alpha_1=1, \, \tau_x=1$ when the
latency periods are controlled by a Poisson event process.
\begin{figure}
\begin{center}
\begin{picture}(300,210)(0,0)
\put(-60,-20){\resizebox{25cm}{!}{\includegraphics{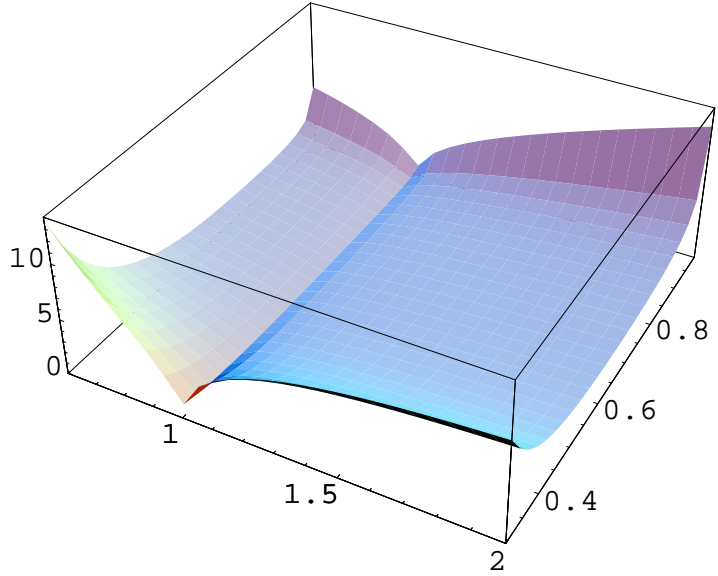}}}
\end{picture}
\end{center} \caption{{\em Approximate information distances $dE_M=\sqrt{E_M}$
(equation (\ref{EM})) in the McKay manifold, measured from
distributions $T_0$ with exponential marginal distribution for
$x$ so $\alpha_1=1$ and $\tau_x=1.$ So the surface represents
distances from an exponential process for periods of latency.}}
 \label{MsqrtEcvxrho}
\end{figure}
The derivation of a distance from distribution $T_0$  is given in~\cite{InfoGeom}, and yields
in terms of $\tau_x$ and $\rho$
\begin{eqnarray}
E_M(\tau_x,\rho)_{|[T_0:\alpha_1=1]} &=& \frac{\left(\rho^2+1\right)^2}{16 \rho ^6}
{\left|\frac{1}{\tau_x^2}-1\right| } \nonumber \\
   &+&   \frac{1}{4}
   \left|\left(1-\frac{1}{\tau_x^2}\right) \left(1-\frac{1}{\rho
   ^2}\right)+3 \log \left(\tau_x^2\right)\right| \nonumber \\
   &+& \left|\psi
   \left(\frac{1}{\tau_x^2}\frac{1}{\rho ^2}-1\right)-\psi
   \left(\frac{1}{\rho ^2}-1\right)\right| \nonumber \\
   & +& \left|\psi
   \left(\frac{1}{\tau_x^2}\right)+\gamma \right|  \label{EM}
\end{eqnarray}
where $\psi(u)=\frac{d\log\Gamma(u)}{du}$ is the digamma function and
$\gamma$ is the Euler gamma constant---with numerical value about $0.577.$
Figure~\ref{MsqrtEcvxrho} shows a plot of $dE_M=\sqrt{E_M(\tau_x,\rho)}$ from
equation (\ref{EM}). This is
an approximation to the Riemannian distance but it represents the
main features of the information
distance of arbitrary latency period distributions  $T_1$ from the curve
of distributions $T_0$  with
$\alpha_1=1, \, \tau_x=1.$

\begin{figure}
\begin{center}
\begin{picture}(300,210)(0,0)
\put(-0,0){\resizebox{10 cm}{!}{\includegraphics{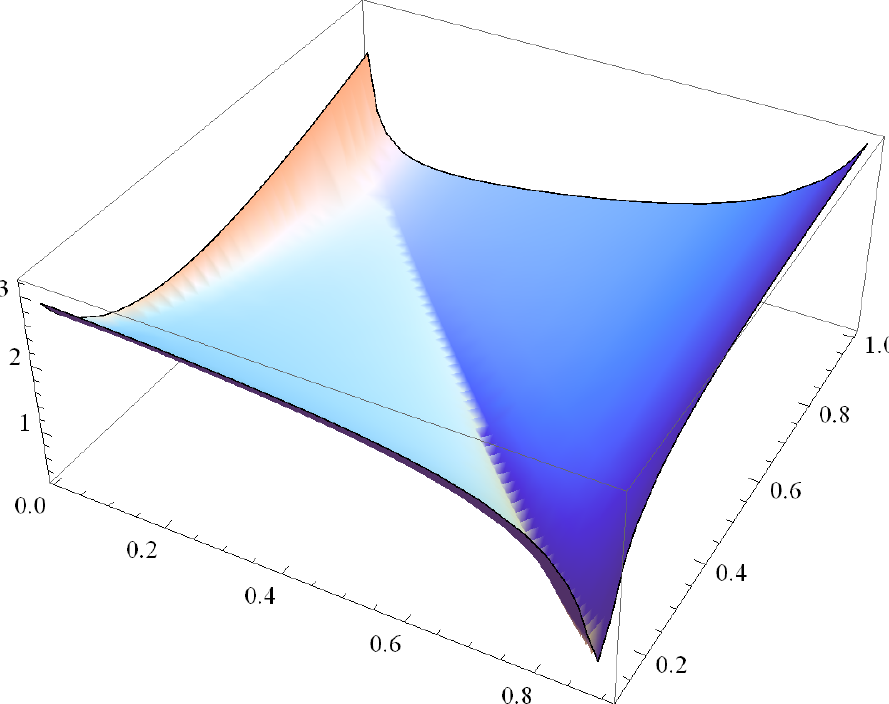}}}
\end{picture}
\end{center} \caption{{\em Approximate information distances $dE_M=\sqrt{E_M}$
(equation (\ref{EMy})) in the McKay manifold, measured from
distributions $T_0$ with exponential marginal
distribution for $y,$ so $\alpha_1+\alpha_2=1$ and $\tau_y=1.$ So the surface represents
distances from a Poisson process for infectivity.}}
 \label{MsqrtEa1a2}
\end{figure}

Repeating the above procedure for the case when $T_0$ has $(\alpha_1+\alpha_2)=1,$
which corresponds to an exponential infectiousness period distribution (and a Poisson
process of infections) we obtain
\begin{eqnarray}
E_M(\alpha_1,\alpha_2)_{|[T_0:\alpha_1+\alpha_2=1]} &=&
 \left|\psi \left(\alpha _2\right)-\psi\left(1-\alpha
   _1\right)\right| \nonumber \\
&+& \frac{1}{4} \left|\frac{\left(2 \alpha _1+\alpha
   _2\right){}^2}{4 \alpha _1}-\frac{1}{2} \left(\alpha
   _1+1\right)\right| .\label{EMy}
\end{eqnarray}
This is plotted in Figure~\ref{MsqrtEa1a2}.
The two graphics, Figures~\ref{MsqrtEcvxrho} and \ref{MsqrtEa1a2}, show how
we can depict the parameters in the joint distribution of periods of latency and
infectiousness as surfaces of distance, measured from the two reference cases
for the evolution of the epidemic
starting from Poisson processes, respectively. On
such surfaces could be represented data on the progress of epidemics under
different intervention schemes, or simulations of such scenarios.

Geodesic curves in Riemannian manifolds give minimal arc length and examples
are given in~\cite{entflow} for manifolds of Weibull, gamma and McKay
bivariate gamma distributions, together with gradient flow curves for entropy.
More details of the information
geometry of uniform, exponential, gamma, Gaussian, and bivariate versions
with applications are provided in \cite{InfoGeom}.

\cite{BL} highlighted aspects
when stochastic features are important and used independent gamma random variables
to represent inhomogeneity of latency and infectiousness periods.
In this paper we have a bivariate inhomogeneous
epidemic process, modeled by correlated gamma distributions and we can use similar methods
to depict and quantify departures from exponential periods of latency and infectiousness.
This shows that the result of ~\cite{BL} for the dependence
of growth rate on variability in the periods of latency and infectiousness in the independent case
persists also in the presence of correlation between the two random variables,
Figures~\ref{r123} and \ref{r3}. Moreover, the information theoretic distance from the two
reference scenarios of exponential distributions of periods latency and infectiousness,
Figures~\ref{MsqrtEcvxrho} and \ref{MsqrtEa1a2}, provide natural quantitative representations
for comparing different parametric data.

 \cite{BL} used independent
gamma distributions for periods of latency and infectiousness, from which the reproduction rate
can be estimated, with applications for example to the SARS outbreak~\cite{who}. Here we have used
a bivariate gamma distribution which allows a corresponding
reproduction rate to be computed. Also, we considered the case when the susceptibility to
infection is not uniform and illustrated with the case when it begins as a gamma
distribution then evolves as the epidemic proceeds. Other models could be used for the
initial distribution of susceptibilities, including asymmetric distributions. A wide range of
such other cases using log-gamma distributions is considered in \cite{evolmodel}
for a similar rate process applied to an evolutionary model when the random variable $a$
represents unfitness (like susceptibility to infection) in a population.


\end{document}